# Irradiation Enhanced paramagnetism on graphene nanoflakes

Andreas Ney[1]*, Pagona Papakonstantinou[2]*, Ajay Kumar[2], Nai-Gui Shang[2], Nianhua Peng[3]

[1] Fakultät für Physik and CeNIDE, Universität Duisburg-Essen, Lotharstr. 1, D-47057 Duisburg, Germany

[2] Nanotechnology and Advanced Materials Research Institute, NAMRI, School of Engineering, University of Ulster, Newtownabbey BT37 0QB, UK

[3] Ion Beam Centre, University of Surrey, Guildford, GU2 7XH, UK.

*Corresponding authors: p.papakonstantinou@ulster.ac.uk; andreas.ney@uni-due.de

**Abstract**

We have studied the magnetization of vertically aligned graphene nanoflakes irradiated with nitrogen ions of 100 KeV energy and doses in the range $10^{11}$- $10^{17}$ ions/cm$^2$. The non-irradiated graphene nanoflakes show a paramagnetic contribution, which is increased progressively by ion irradiation at low doses up to $10^{15}$ /cm$^2$. However, further increase on implantation dose reduces the magnetic moment which coincides with the onset of amorphization as verified by both Raman and X-ray photoelectron spectroscopic data. Overall, our results demonstrate the absence of ferromagnetism on either implanted or unimplanted samples from room temperature down to a temperature of 5K

Ferromagnetism from sp electrons is not new and has been observed in many allotropes of carbon [1-4]. However, in all these studies the magnetization signals observed were very small ($M_s \sim 0.1$ emu/g)..

So far the number of experimental works on room temperature graphene ferromagnetism even though is quite limited, has produced some conflicting and inconsistent results. For example, small magnetic signals ($M_s = 0.1$-1 emu/g) were detected in ensembles of graphene oxide sheets reduced by a combination of hydrazine and annealing[5]. However at an annealing temperature of 800 °C the saturation magnetization exhibited an abrupt negative trend, which is in contrast to the expectation of observing increased magnetization for higher reduction temperatures. Although small magnetic signals ($M_s < 0.7$ emu/g) at low temperatures were observed in graphene sheets of small size possessing a large amount of edge defects, produced by arc evaporation of graphite[6] in a hydrogen atmosphere, no ferromagnetic signals were observed in graphene obtained by extensive sonic exfoliation[7], a method which also produces graphene of small sizes (< 50nm). Thus, further experiments are necessary to clarify the picture.

Recently we have successfully grown self assembled vertically aligned multilayer graphene nanoflake films (MGNFs), which terminate on a few graphene layers (~2 nm thick edges) using microwave plasma enhanced chemical vapour deposition[8]. These graphene sheets are grown on a Si substrate without the use of any catalysts and therefore present an ideal system for exploring magnetism due to the absence of metal growth catalysts. Controllable introduction of defects by irradiation followed by magnetic measurements, can be particularly useful to explore if irradiation can essentially induce magnetic ordering in graphene nanoflakes. By varying the ion irradiation conditions we have found that either unimplanted or implanted

MGNFs show only paramagnetic contributions with absence of ferromagnetic ordering even at 5K.

Figures 1(a) and (b) are lower and higher magnification scanning electron microscopy (SEM) images showing the top view of a dense carpet of quasi vertically oriented curled graphene nanoflakes. The resulting coating grown on n doped Si substrates is approximately 15 µm high (Fig. 1(c)). The relatively high thickness of the coating, warrants that the implanted ions do not reach the substrate. All MGNFs were produced under exactly the same controllable and reproducible deposition conditions. To account for any differences on the magnetic behavior induced by variation on the surface morphology, magnetic measurements were repeated on two samples produced under identical conditions. The implantation was performed at room temperature with 100 KeV N-ions at doses of $10^{12}$, $10^{13}$, $10^{14}$, $10^{15}$, $10^{16}$ and $10^{17}$ ions/cm$^2$ at the Ion Beam Centre of Surrey University. The magnetic measurements were conducted using a commercial superconducting interference device (SQUID) magnetometer (Quantum Design MPMS XL) applying the magnetic field in the film plane. M(H)-curves were measured at 300 K and 5 K respectively. In addition temperature dependent magnetisation curves were measured at 10 mT, while warming the sample after either cooling in a field of 4 T (FC) or at zero field (ZFC). The diamagnetic contribution of the Si substrate was measured separately by determining the ratio of the magnetic background between 300 K and 5 K, in order to eliminate the small temperature dependent van-Vleck-type paramagnetic contributions[9]. In general the net-diamagnetic background was generally found to be by ~10% larger at 5 K than at 300 K. This ratio was used to calculate the net-diamagnetic background at 5 K from the 300 K measurements at magnetic fields between 1 T and 4 T of each sample. This assures reliable determination of the shape and curvature of the M(H) curve at 5 K. Potential ferromagnetic contamination by metal tweezers handling or text-marker writings on the back of the substrate was carefully avoided. Proton-induced ray emission (PIXE) spectra with a micro-beam measured before and after irradiation of the samples revealed no detectable metal signals.

Figure 2 shows the typical Raman spectra of pristine and irradiated MGNFs, excited by 514 nm Ar$^+$ ion laser using a back-scattering configuration at room temperature. For the pristine MGNFs, there are four peaks located at 1361, 1588, 1628 and 2713 cm$^{-1}$, which are assigned to the D, G, D´ and G´ bands of crystalline graphitic materials, respectively. The G band corresponds to the in-plane stretching vibration of sp$^2$ carbon atoms in a graphitic 2D hexagonal lattice, and the D band is used as evidence of the disruption of the aromatic system of π-electrons in the graphitic framework. The G´ band located at ~2713 cm$^{-1}$, originates from a double-resonance process[10]. An increase in the number of defects results in an increase of the *D* peak intensity and a simultaneous drop in the intensity of the G´ peak [11]. In addition, the D′ (~1628 cm$^{-1}$) peak is also a lattice disorder induced band from the double resonance Raman process due to intravalley scattering. Upon subjection of MGNFs to N$^+$ ion irradiation doses of 10$^{12}$, 10$^{13}$, and 10$^{14}$ ions/cm$^2$, the intensity of both D and D´ bands is observed to increase gradually whereas both G and G´ bands weaken. Usually, the ratio ($I_D/I_G$) of integrated intensities of the D band to the G band is used to characterize the crystalline quality of graphitic materials: the larger the more defective. The ratio displayed in the Fig. 2 increases from 0.347 to 1.563 with the increase of ion doses from 10$^{12}$ to 10$^{14}$ ions/cm$^2$, revealing that the ion irradiation creates a number of structural defects in MGNFs and the defect density increases with the ion dose. At an ion dose of 10$^{15}$ ions/cm$^2$, the G´ band disappears and there are only two broader D and G bands, which is a typical feature of nanometer sized graphitic materials. This manifests that the structure of MGNFs is changed from the crystalline graphitic materials[12] into nanocrystalline graphite under the ion irradiation dose of $10^{15}$ ions/cm$^2$. Further increase of the ion dose from $10^{16}$ to $10^{17}$ ions/cm$^2$ causes both D and G bands to become broader and gradually overlapped in parts. The whole Raman spectrum for the ion doses of $10^{17}$ ions/cm$^2$ presents a typical feature of amorphous carbon or carbon nitride, where either carbon or nitrogen atoms are completely in a disordered state. However, no carbon nitride related peaks are observed. This clearly points out that the irradiated MGNFs are transformed from a nanographitic to an amorphous material after the high dose of irradiation. Here, high energetic N$^+$ ions implant into the matrix of MGNFs and migrate the carbon atoms in the lattice producing point defects. With the increasing of ion dose, a large quantity of point defects is produced and accumulated, finally leading to disordering and amorphization. Such point defects could be either C vacancy or N interstitials/substitutes, which possibly form carbon nitride. Therefore, N$^+$ ion irradiations induce structural defects in MGNFs, and as a result their structure evolves gradually from the crystalline graphene nanoflake to a nanographitic structure and then finally to amorphous carbon or carbon nitride formation.

Further corroborative evidence for amorphisation at high implantation doses is provided by X-ray photoelectron spectroscopy (XPS) data. High resolution XPS analysis was carried out using a SCIENTA ECSA 300 equipped with a monochromatic Al K$\alpha$ (h$\nu$ = 1486.6 eV) X-ray source at Daresbury Laboratory. The width of C1s core level spectra increased slightly from 0.61 to 0.67 for irradiation doses from 0 up to $10^{14}$ ions/cm$^2$. However further increase in the range $10^{15}$ to $10^{17}$ ions/cm$^2$ caused an amplified broadening of the C1s ranging from 0.94 to 1.3 indicating augmented amorphisation (not shown). Since the beam irradiates the surface of MGNF coating perpendicularly, the ion beam is basically parallel to the main portion of the graphene nanoflakes, which have a quasi vertical alignment. Typical SEM images of the pristine and implanted MGNFs with a N$^+$ dose of 2 × $10^{17}$ ions cm$^{-2}$ (Figure 1d) revealed that the irradiated MGNFs preserved their 3D network structure, with the characteristic branching morphology. Both XPS (penetration depth <10 nm) and Raman spectroscopy with the excitation of 514 nm laser (penetration depth > 100 nm) confirm that implantation damage was produced under the irradiation of N ions. SRIM simulation shows that 100 keV N ions have an approximated projected range of 340 nm based on a density of 1.5 g/cm$^3$ (estimated through measurements of film mass and volume). Therefore the implantation-induced disorder extends from the surface to a depth of about 340 nm. The top part of the graphene nanoflakes is transformed into highly disordered structure (amorphous carbon), when the dose is sufficiently high, while the bottom part still retains its highly crystalline graphitic structure.

Figure 3 summarizes the results of the magnetic measurements of a series of N-ion implanted graphene nanoflakes using SQUID magnetometry. Figure 3(a) shows exemplarily the M(H) curves recorded at 300 K and 5 K of the sample implanted with a N-dose of $10^{13}$/cm$^2$. The M(H) curves are anhysteretic within the artifact level of the SQUID magnetometer of 0.2 μemu[13]. The paramagnetic behavior of this sample is also confirmed by M(T) measurements, which do not show a significant separation between FC and ZFC (not shown). In addition, the M(H) curve at 5 K can be well-reproduced by a Brillouin-type function using *J*=*S*=1 and *g*=2 corresponding to two magnetically coupled unpaired electrons. Figure 3(b) summarizes the essence of the N-implantation series from unimplanted (pristine) graphene nanoflakes up to a N-dose of $10^{17}$/cm$^2$. Shown are the magnetization values at 4 T measured at 5 K (left axis) and 300 K (right axis) normalized by the sample area. First, one has to note that the magnetization at 300 K is in general by a factor of 20 to 60 smaller than at 5 K. Moreover, all magnetization values at 300 K are between 0.2 and 0.5 μemu and therefore of the same order of the typical artifact level of the SQUID[13]. Therefore, at no dose from $10^{13}$ to $10^{17}$/cm$^2$ any significant evidence for ferromagnetic order at 300 K can be found. This is also corroborated by FC/ZFC data (not shown). These observations are in disagreement with the reports of room temperature induced ferromagnetism in irradiated graphite or room- temperature ferromagnetism in graphene like materials[1-6]. Second, the unimplanted graphene nanoflakes show a small (~18 μemu) but finite paramagnetic signal at 5 K. The origin of this paramagnetism can either be related to intrinsic contributions from the nonbonding π-electron states localized at the graphene layer edges (carbon dangling bonds at the edge planes) or to the procedure of determining the net-diamagnetic background of the substrate (see above). On the other hand, this result rules out the possibility of ferromagnetic contamination of the C used for this batch of samples. Despite the absence of ferromagnetism, the ion implanted samples exhibit noticeable low temperature paramagnetism. The disorder created by the ion bombardment enhances this paramagnetic contribution. The magnetization at 5 K and 4 T initially increases with increasing N-dose reaching its maximum at $10^{15}/cm^2$ and decreases at higher doses which can be taken as an indication of increasing amorphization in accordance with the Raman spectra. The agreement between SQUID and Raman results proves that our results are reliable and do not originate from impurities as was confirmed by our PIXE results, which do not show any traces of metal contamination. This evidences, that the observed paramagnetism is merely related to structural effects induced by ion irradiation.

To extract more quantitative information from the magnetic data, we subtract the paramagnetism of the pristine graphene sample from the other magnetic data. Thus, we can attribute any additional paramagnetic signal to the implanted N-ions. From the area-dose one can easily calculate the magnetic moment per implanted ion. The outcome is displayed as a function of N-dose in Figure 3(c) on a log-log scale. Remarkably, the effective magnetic moment per implanted N-ion is as high as $400\mu_B$ at a dose of $10^{13}$/cm$^2$ and it strongly decreases down to $0.1\mu_B$ at a dose of $10^{17}$/cm$^2$. Considering the magnetic moment of $2\mu_B$ extracted from the M(H) curvature, each N-ion produces 200 uncoupled paramagnetic moments of $2\mu_B$ each. A coupled supermoment of $400\mu_B$ would be inconsistent with such a weak curvature of the M(H) data at 5 K. We tentatively ascribe the origin of these paramagnetic moments of $2\mu_B$ to the creation of defects caused by the implantation process. A possible defect can be a broken bond where two unpaired electrons are magnetically coupled. However, based on the magnetic data no unambiguous identification of the nature of the defect is possible. On the other hand, it is obvious that the creation of such defects is most efficient at low doses, i.e. at a dose of $10^{13}$/cm$^2$ each N-ion creates about 200 of such defects. Considering the energy of the impinging N-ions of 100 keV and a film density of 1.5 g/cm$^3$ and using a modified Kinchin-Pease model with the threshold energy of displacement set at 28 eV, the SRIM calculation predicts that each ion can create up to 300 defects, of which 2/3 have to be stable and paramagnetic. This estimation demonstrates that such an effective magnetic moment of 400 $\mu_B$ per implanted N-ion is not unreasonably large. Of course, with increasing N-dose many defects have already been created by the first ions so that the efficiency of creating further stable defects may drastically go down. In addition, an increasing amorphization of the graphene nanoflakes may also destabilize the paramagnetic nature of the defects.

A. N. gratefully acknowledges financial support from the German Research Foundation (DFG) within the Heisenberg-Programm. The work was supported by the EPSRC Funded Facility access to Ion beam Centre at Surrey University and the University of Ulster (PhD VCRS studentship to A. K.).

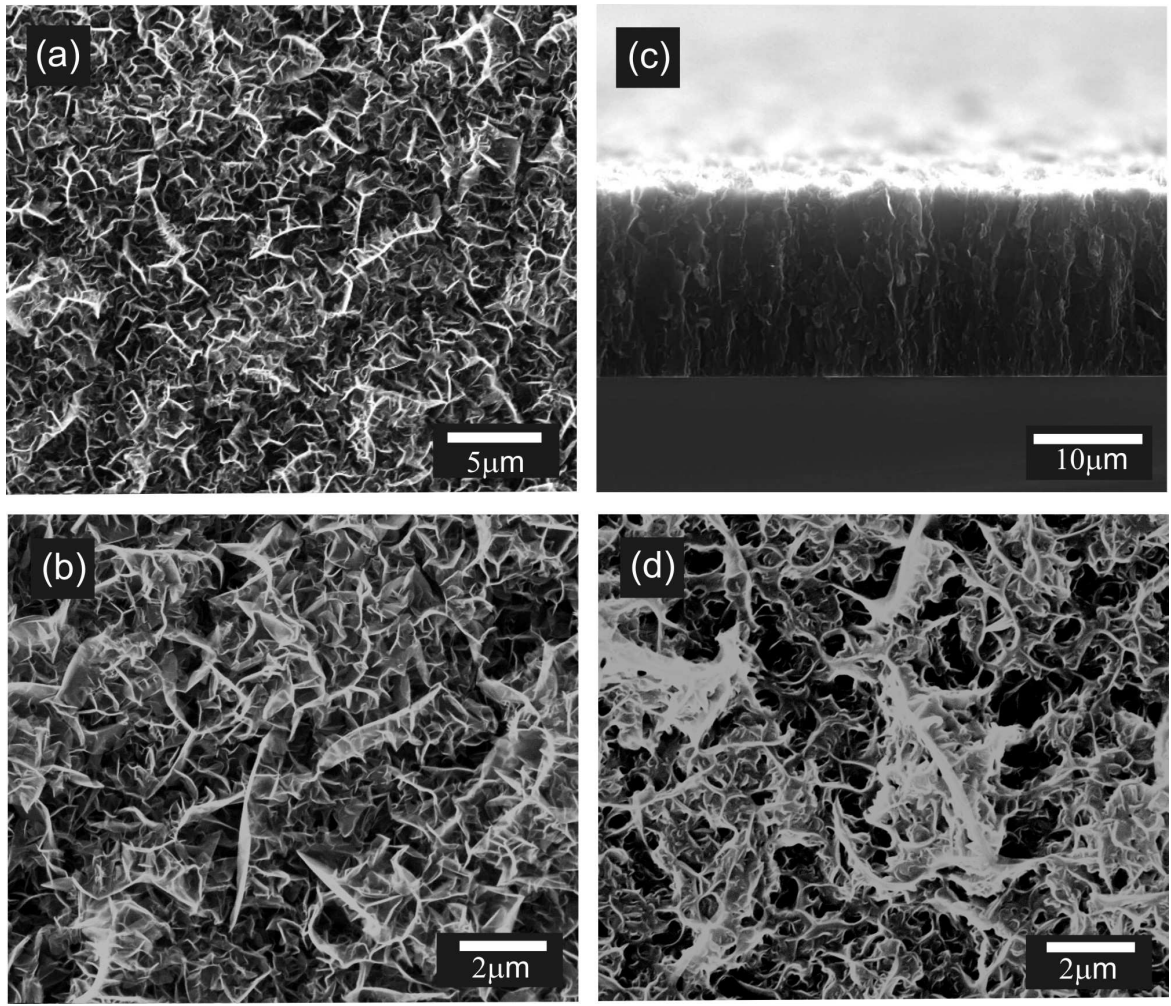

**Figure 1**

(a) (b) Low and higher magnification SEM micrographs of MGNFs displaying a crumpling and branching morphology. Each graphene flake has a wedge like structure with a 2nm or less thick edge and a progressively thicker base; (c) The cross-section demonstrates the vertical orientation of the MGNFs; (d) SEM micrograph of MGNFs irradiated with an ion dose of $10^{17}$ ions/cm$^2$

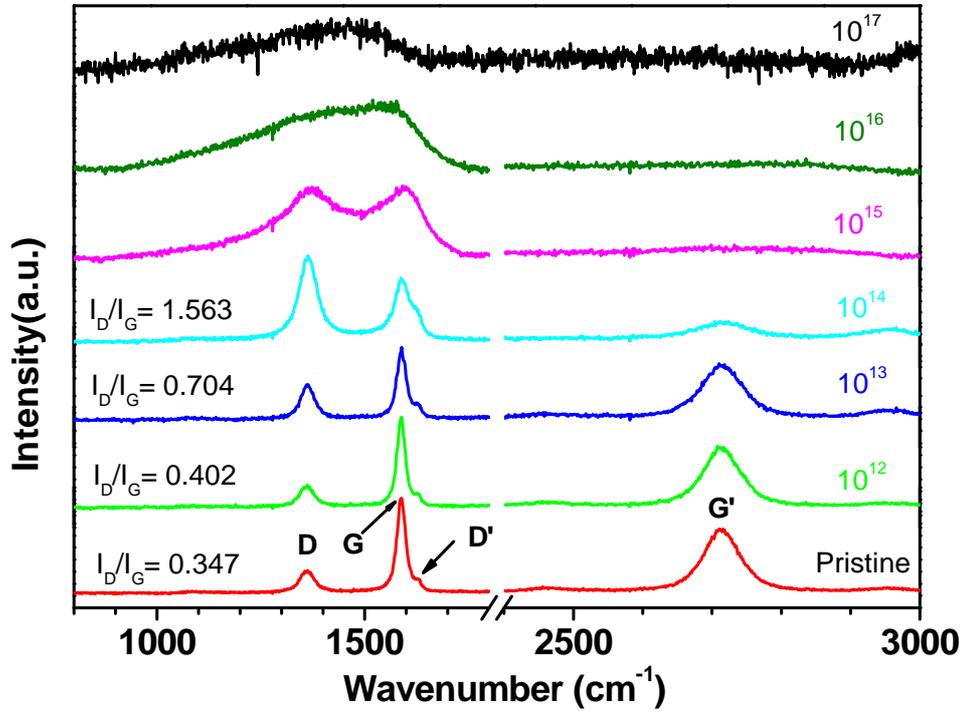

**Figure 2**

Raman spectra of multilayer graphene nanoflakes implanted with N-doses of 0 (pristine) to $10^{17}$ ions/ cm$^2$. The non distinguishable D and G peaks and absence of G' peak at $10^{15}$ ions/ cm$^2$ indicate augmented amorphisation

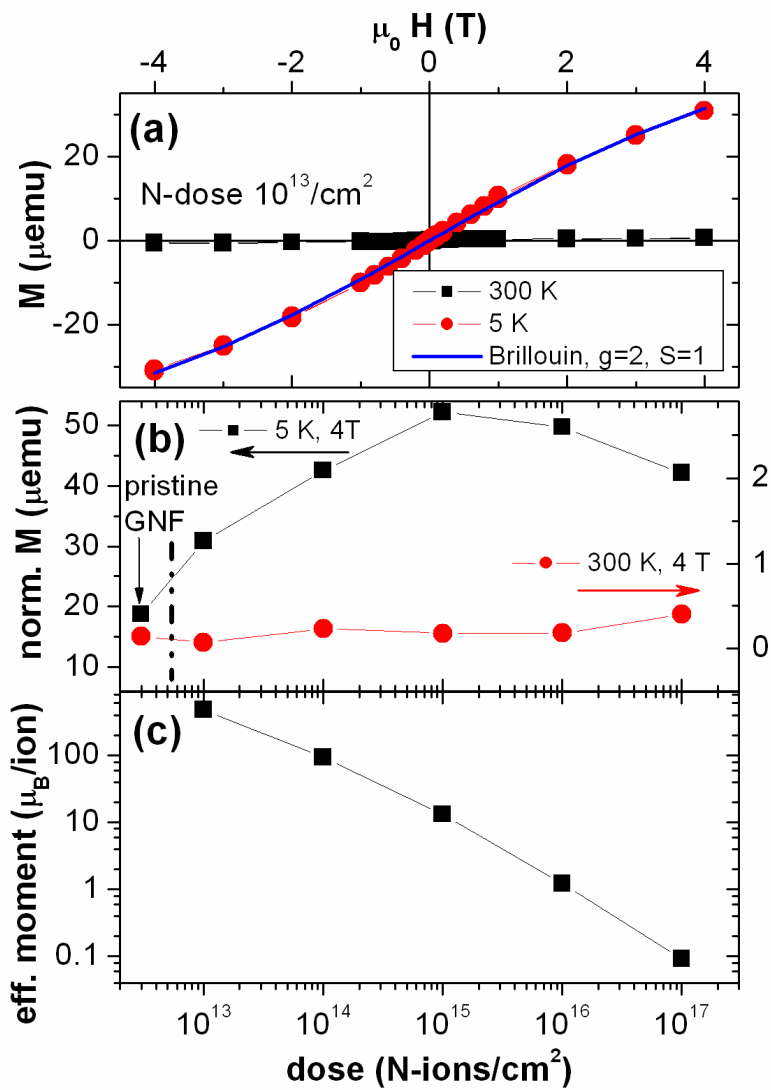

**Figure 3**. Magnetic response of nitrogen implanted graphene nanoflakes with doses of 0 to $10^{17}$ ions/cm$^2$ (a) Magnetic moment M as a function of field, H at temperatures of 5 and 300 K for a MGNF implanted with $10^{13}$ ions/cm$^2$. Symbols represent the measurements and the line represents the fitted Brillouin curve at 5K using J=S=1 and g=2 values. (b) Magnetization values normalized by the sample area as a function of implantation dose measured at temperatures of 5 and 300 K. (c) Effective magnetic moment per implanted N-ion as a function of implantation dose.